\RequirePackage{fix-cm}
\documentclass[smallextended]{svjour3}       
\smartqed  
\usepackage{graphicx}
\usepackage{amsmath,amsfonts,bm}
\usepackage[breaklinks=true]{hyperref}
\usepackage{breakurl}
\begin{document}

\title{Improving genetic risk prediction by leveraging pleiotropy
}
\subtitle{}


\author{Cong Li         \and
        Can Yang \and
        Joel Gelernter \and
        Hongyu Zhao 
}


\institute{Cong Li \at
              Program in Computational Biology and Bioinformatics, Yale University, New Haven, Connecticut 06520, USA\\
              \email{cong.li@yale.edu}
           \and
	Can Yang \at
              Department of Biostatistics, Yale School of Public Health, Yale University, New Haven, Connecticut 06520, USA \\
              Department of Psychiatry, Yale University School of Medicine, New Haven, CT, USA\\
               \email{can.yang@yale.edu}           
           \and
           Joel Gelernter \at
           Department of Psychiatry, Yale University School of Medicine, New Haven, CT, USA\\
           VA CT Healthcare Center, Departments of Genetics and Neurobiology, Yale Univ. School of Medicine, West Haven, Connecticut 06516, USA\\
           \email{joel.gelernter@yale.edu}
           \and
           Hongyu Zhao (Corresponding author)\at
           Department of Biostatistics, Yale School of Public Health, Yale University, New Haven, Connecticut 06520, USA\\
           Program in Computational Biology and Bioinformatics, Yale University, New Haven, Connecticut 06520, USA\\
           \email{hongyu.zhao@yale.edu}
}

\date{Received: date / Accepted: date}

\maketitle

\begin{abstract}

An important task of human genetics studies is to accurately predict disease risks in individuals based on genetic markers, which allows for identifying individuals at high disease risks, and facilitating their disease treatment and prevention. Although hundreds of genome-wide association studies (GWAS) have been conducted on many complex human traits in recent years, there has been only limited success in translating these GWAS data into clinically useful risk prediction models. The predictive capability of GWAS data is largely bottlenecked by the available training sample size due to the presence of numerous variants carrying only small to modest effects. Recent studies have shown that different human traits may share common genetic bases. Therefore, an attractive strategy to increase the training sample size and hence improve the prediction accuracy is to integrate data of genetically correlated phenotypes. Yet the utility of genetic correlation in risk prediction has not been explored in the literature. In this paper, we analyzed GWAS data for bipolar and related disorders (BARD) and schizophrenia (SZ) with a bivariate ridge regression method, and found that jointly predicting the two phenotypes could substantially increase prediction accuracy as measured by the AUC (area under the \emph{receiver operating characteristic} curve). We also found similar prediction accuracy improvements when we jointly analyzed GWAS data for Crohn's disease (CD) and ulcerative colitis (UC). The empirical observations were substantiated through our comprehensive simulation studies, suggesting that a gain in prediction accuracy can be obtained by combining phenotypes with relatively high genetic correlations. Through both real data and simulation studies, we demonstrated pleiotropy as a valuable asset that opens up a new opportunity to improve genetic risk prediction in the future.

\end{abstract}

\section{Introduction}
\label{intro}

Predicting disease risks in individuals based on genetic markers, which is usually referred to as ``genetic risk prediction'', is an important task in human genetics studies \cite{collins2001implications}. Although hundreds of genome-wide association studies (GWAS) have been conducted and many thousands of genomic regions have been implicated for various human traits in recent years, these findings have not been translated into clinically useful risk prediction models based on genetic markers, which limits their potential impact on personalized disease prevention and treatment. Most genetic risk prediction models, including those used by direct-to-customer genetic testing companies, are constructed based on a few significant (usually validated) single nucleotide polymorphisms (SNPs). However, such SNPs typically account for only a small fraction of the total heritability and thus cannot provide satisfactory prediction accuracy \cite{manolio2009finding}. A large fraction of the genetic variances is not accounted by such genetic risk predictions because they were ``missing'' in the SNPwise GWAS results \cite{maher2008case}. Several hypotheses have been put forward to explain the ``missing heritability" \cite{eichler2010missing}, including epistasis \cite{clarke2010gwas,huebinger2010pathway}, undetected CNVs \cite{forer2010conan}, and rare variants \cite{dickson2010rare}, among others. However, it has also been suggested that much of the missing heritability is in fact hidden among the numerous common genetic variants carrying only small to modest effects \cite{gibson2010hints,visscher2012five,park2010estimation}. Through a polygenic risk-score analysis on GWAS data of schizophrenia (SZ) and bipolar disorder (BP), Purcell et al \cite{purcell2009common} inferred that the genetic risk factors for SZ and BP may involve thousands of common SNPs of very individually small effect sizes. Using a linear mixed model, Yang et al \cite{yang2010common} estimated that whole-genome common SNPs could explain about 45\% of the human height variance. Lee et al \cite{lee2011estimating} extended this method to complex human diseases and also found that a substantial fraction of the phenotypic variances could be explained by genotyped common SNPs. In this manuscript, we refer to the proportion of phenotypic variance explained by all genotyped common SNPs as the ``chip heritability'' \cite{zhou2013polygenic}, denoted as $h^2$. \\

These results suggest that, instead of building genetic prediction models with only a few significant variants, a more sensible approach would be to impose less stringent criteria on variant selection or even to build prediction models using whole genome variants \cite{de2010predicting}. In this way, more variants with weak effects can be used though at the expense of also including many variants that do not affect the phenotype one way or the other, or even act in the opposite effect direction. Following Makowsky et al \cite{makowsky2011beyond}, we refer to this class of methods as Whole Genome Prediction (WGP) methods. However, there is still a significant gap between the prediction accuracy achieved and the genetic variance accounted by all common SNPs, even for the WGP methods \cite{makowsky2011beyond}. These authors also investigated several parameters that affect prediction accuracy and showed that a substantial gain may be achieved by increasing the training sample size, and using training samples that are more related to the testing samples. Wray et al \cite{wray2013pitfalls} also pointed out limited sample size as a bottleneck to achieve accurate genetic risk predictions.\\

These studies suggest that, in the presence of numerous weak-effect variants for most complex traits , one of the most effective ways to develop a more accurate risk prediction model is to increase the size of the training sample. But subject recruitment may be difficult and expensive. Alternatively, if another phenotype shares some genetic variants that affect the primary phenotype of interest, it might be possible to incorporate the available data for this correlated phenotype in risk model developments. In this way, we can increase the effective training sample size with the caveat that the additional samples are not directly related to the primary phenotype of interest. Appropriate statistical methods are needed to jointly analyze these distinct yet related data sets.\\

In fact, there is accumulating evidence suggesting that different complex human traits are genetically correlated, i.e. multiple traits share common genetic bases, which is also formally known as ``pleiotropy". In a systematic analysis of the open-access NHGRI catalog, ~17\% of the trait-associated genes and ~5\% of the trait-associated SNPs showed pleiotropic effects \cite{sivakumaran2011abundant}. Vattikuti et al \cite{vattikuti2012heritability} used a bivariate linear mixed model to analyze the Atherosclerosis Risk in Communities GWAS and found significant genetic correlations between several metabolic syndrome traits, including body-mass index, waist-to-hip ratio, systolic blood pressure, fasting glucose, fasting insulin, fasting trigylcerides, and fasting high-density lipoprotein. Lee et al \cite{lee2012estimation} extended this bivariate linear mixed model so that it could deal with binary traits, e.g. presence or absence of a disease.  Andreassen et al \cite{andreassen2013improved} applied a ``pleiotropic enrichment" method on GWAS data of schizophrenia and cardiovascular-disease and showed that the power to detect schizophrenia-associated common variants can be improved by exploiting the pleiotropy between these two phenotypes. More recently, a study on genome-wide SNP data for five psychiatric disorders in 33,332 cases and 27,888 controls identified four significant loci ($P < 5 \times 10^{-8}$) affecting multiple disorders, including two genes encoding two L-type voltage-gated calcium channel subunits, \emph{CACNA1C} and \emph{CACNB2} \cite{smoller2013identification}. Results from the large scale Collaborative Oncological Gene-environment Study also highlighted the existence of ``carcinogenic pleiotropy'', i.e. the overlap between loci that confer genetic susceptibility to multiple types of tumor \cite{sakoda2013turning}. \\

All these findings are exciting because they imply that genetic correlation are prevalent among complex human diseases and hence leveraging the genetic correlations between phenotypes might be a promising strategy to improve genetic risk prediction in the future. Although genetic correlations have been extensively studied for association analyses \cite{korte2012mixed,huang2011prime}, little attention has been paid to their utility in genetic risk prediction. In this paper, we propose to use a bivariate ridge regression method to leverage the genetic correlation between two diseases in genetic risk prediction. We analyzed real GWAS data sets for two pairs of related common diseases. We performed a comprehensive simulation study on the utility of genetic correlation by investigating the gain of prediction accuracy as a function of the strength of genetic correlation between two traits. We also examined the effects of several other parameters such as the ``chip heritability'' $h^2$, the training sample size and the number/proportion of causal SNPs.  \\

\section{Methods}\label{sec_methods}

In this study, four dbGap data sets were analyzed. We didn't obtain the consent from the participants because we downloaded the data indirectly from database. Moreover, the data were analyzed anonymously. For two of the dbGap datasets, we obtained the approval of the institutional review board. For the other two data sets, institutional review board approval is not required to access the data sets.

\subsection*{Data Description}

We downloaded a GWAS data set of bipolar and related disorders (BARD) and a GWAS data set of schizophrenia (SZ) from the dbGaP database (\burlalt{{https://dbgap.ncbi.nlm.nih.gov}}{https://dbgap.ncbi.nlm.nih.gov}). All the subjects were genotyped on the Affymetrix Genome-Wide Human SNP Array 6.0 platform. See \burlalt{{http://www.ncbi.nlm.nih.gov/projects/gap/cgi-bin/study.cgi?study\_id=phs000017.v3.p1}}{http://www.ncbi.nlm.nih.gov/projects/gap/cgi-bin/study.cgi?study_id=phs000017.v3.p1} and \burlalt{{http://www.ncbi.nlm.nih.gov/projects/gap/cgi-bin/study.cgi?study\_id=phs000021.v3.p2}}{http://www.ncbi.nlm.nih.gov/projects/gap/cgi-bin/study.cgi?study_id=phs000021.v3.p2} for more details. We merged the two data sets after removing redundant subjects and collected the genotype data for subjects with only European ancestry. The merged data set consists of 653 affected subjects for BARD, 1170 affected subjects for SZ and 1403 controls. SNPs with minor allele frequency (MAF) smaller than 0.05 and missing rate greater than 0.01 were excluded. The remaining missing genotypes were randomly drawn from binomial distributions based on the allele frequencies at the given loci. SNPs that failed the Hardy-Weinberg Equilibrium test ($P < 0.0001$) in either BARD, SZ or control group were also excluded. We also performed linkage-disequilibrium pruning so that every pair of SNPs within a 50-SNP window had an R-squared value no greater than 0.8. After these procedures,  298,604 SNPs remained. 

For the second pair of diseases, we downloaded a GWAS data set of Crohn's disease (CD) and a GWAS data set of ulcerative colitis (UC).  The subjects in the CD data set were genotyped on the ILLUMINA HumanHap300v1.1 platform. See \burlalt{{http://www.ncbi.nlm.nih.gov/projects/gap/cgi-bin/study.cgi?study\_id=phs000130.v1.p1}}{http://www.ncbi.nlm.nih.gov/projects/gap/cgi-bin/study.cgi?study_id=phs000130.v1.p1} for more details. UC subjects were genotyped on either the ILLUMINA HumanHap300v2 or the HumanHap550v3 platform. See \burlalt{{http://www.ncbi.nlm.nih.gov/projects/gap/cgi-bin/study.cgi?study\_id=phs000345.v1.p1}}{http://www.ncbi.nlm.nih.gov/projects/gap/cgi-bin/study.cgi?study_id=phs000345.v1.p1} for more details. Since the SNPs on the HumanHap300v1.1 platform are a subset of the SNPs on the other two platforms, we retained only the HumanHap300v1.1 SNPs in all the data sets for further analysis. In the CD data set, only subjects with non-Jewish, European ancestry were included in analysis, which include of 515 unaffected subjects and 513 affected subjects. The UC data set consists of 902 non-Hispanic, European ancestry, affected individuals. To have a genetically matched control data set for UC, we downloaded all the Caucasian subjects that have genotypes from the HumanHap300v1.1 SNPs from the ILLUMINA iControl database (\burlalt{{http://www.illumina.com/science/icontroldb.ilmn}}{http://www.illumina.com/science/icontroldb.ilmn}). We performed principal component analysis on the iControl samples and the UC samples. For each UC sample, we selected an iControl sample with the smallest Euclidean distance between their first 20 principal components. SNPs that were overlapped between all these data sets were used for analysis after excluding SNPs with minor allele frequency smaller than 0.05 and missing rate greater than 0.01. The remaining missing genotypes were randomly drawn from binomial distributions based on the allele frequencies at the given loci. SNPs that failed the Hardy-Weinberg Equilibrium test ($P < 0.0001$) in either data set were also excluded. We also performed linkage-disequilibrium pruning so that every pair of SNPs within a 50-SNP window had an R-squared value no greater than 0.8. After these procedures,  241,649 SNPs remained. \\

\subsection*{Bivariate Ridge Regression}

To take advantage of the pleiotropy, we used a bivariate ridge regression method to predict jointly the status for two diseases. In fact, the bivariate (or multivariate) ridge regression was motivated as a generalization of univariate ridge regression to account for the ``across regression'' correlations when multiple linear regressions are simultaneously considered \cite{brown1980adaptive,haitovsky1987multivariate}. Similar to the relationship between the univariate ridge regression and the univariate linear mixed model, the bivariate ridge regression is also closely related to the bivariate linear mixed model, which has been implemented to estimate the genetic correlation between phenotypes recently \cite{lee2012estimation}. Here we highlight this connection while we go through the formulation of bivariate ridge regression. \\

Consider the following bivariate linear mixed model\cite{thompson1973estimation}: 
\begin{equation}
 \bf{y^{(1)}} = X^{(1)}\beta^{(1)} + g^{(1)} + e^{(1)}, \nonumber
\end{equation}
\begin{equation}
 \bf{y^{(2)}} = X^{(2)}\beta^{(2)} + g^{(2)} + e^{(2)}
\end{equation}
where $\bf{y^{(1)}}$ and $\bf{y^{(2)}}$ are the two vectors of two phenotypes measured on $n_1$ and $n_2$ individuals, respectively. In this study, $\bf{X}^{(1)}$ and $\bf{X}^{(2)}$ consist of only a column of ones respectively, indicating that only the intercept is included as fixed effect. In this case, $\beta^{(1)}$ and $\beta^{(2)}$ are simply the means of the two phenotypes. In the rest of this paper, we assume that $\bf{y^{(1)}}$ and $\bf{y^{(2)}}$ are already mean-subtracted for convenience. $\bf{g}^{(1)}$ and $\bf{g}^{(2)}$ are the genetic values for the two phenotypes. $\bf{e}^{(1)} $ and $\bf{e}^{(2)}$ are the residuals. The genetic values $\bf{g}^{(1)}$ and $\bf{g}^{(2)}$ are treated as random effects: 
\begin{equation}
\bf{g}^{(1)} = \bf{G}^{(1)}\bf{u}^{(1)}, \,\, \bf{g}^{(2)} = \bf{G}^{(2)}\bf{u}^{(2)}
\end{equation}
\begin{equation}
\left[ \begin{array}{cc} \bf{u}_j^{(1)} \\ \bf{u}_j^{(2)} \end{array} \right] \sim \mathcal{N}(\bf{0}, \bf{\Sigma_g})
\end{equation}
\begin{equation}
\bf{\Sigma_g} = \left[ \begin{array}{cc}
\sigma_{g_1}^2 & \rho_{g}\sigma_{g_1}\sigma_{g_2} \\
\rho_{g}\sigma_{g_1}\sigma_{g_2} & \sigma_{g_2}^2 \end{array} \right]
\end{equation}
where $\bf{G}^{(t)}$ is the standardized genotype matrix for a total of $p$ SNPs of the training samples for the $t$-th phenotype. Specifically, let $B$ and $b$ be the two alleles at the $j$-th locus and $f_j$ be the frequency of the $B$ allele, then $\bf{G}^{(t)}_{ij}$ takes a value of $-2f_j/\sqrt{2f_j(1-f_j)p}$, $(1-2f_j)/\sqrt{2f_j(1-f_j)p}$ or $2(1-f_j)/\sqrt{2f_j(1-f_j)p}$ if the genotype of the $i$-th individual at the $j$-th locus is $bb$, $Bb$ or $BB$, respectively. $\bf{u}_j^{(1)}$ and $\bf{u}_j^{(2)}$ are the random effects of the $j$-th locus for the two phenotypes and $\rho_g$ measures the strength of genetic correlation between the two phenotypes. We also assume the residuals follow a multivariate normal distribution:

\begin{equation}
\left[ \begin{array}{cc} \mathbf{e^{(1)}} \\ \mathbf{e^{(2)}} \end{array} \right] \sim \mathcal{N}(\bf{0}, \bf{\Sigma_e})
\end{equation}
\begin{equation}
\bf{\Sigma_e} = \left[ \begin{array}{cc} \sigma^2_{e_1}\mathbf{I_{n_1}} & 0 \\ 0 & \sigma^2_{e_2}\mathbf{I_{n_2}} \end{array} \right]. 
\end{equation}

In Lee et al.\cite{lee2012estimation}, the random effects $\bf{u}^{(1)}$, $\bf{u}^{(2)}$, $\bf{e}^{(1)}$ and $\bf{e}^{(2)}$ were integrated out and the average information restricted maximum likelihood (AI-REML) algorithm \cite{gilmour1995average} was used to estimate the variance parameters $\sigma_{g_1}^2$, $\sigma_{g_2}^2$, $\rho_g$, $\sigma_{e_1}^2$, $\sigma_{e_2}^2$. $\rho_g$ measures the genetic correlation between the two phenotypes whereas the other four variance parameters can be used to calculate the ``chip'' heritability or ``variance explained by SNPs'' for the two phenotypes. However, our focus here is not to estimate the ``chip heritability'' or the genetic correlation but to predict the phenotypes. Therefore we are more interested in estimating the random effects: $\bf{u}^{(1)}$ and $\bf{u}^{(2)}$. Given a certain set of variance parameters $\sigma_{g_1}^2$, $\sigma_{g_2}^2$, $\rho_g$, $\sigma_{e_1}^2$ and $\sigma_{e_2}^2$, the posterior means of the random effects $\bf{u}^{(1)}$ and $\bf{u}^{(2)}$ can be written out as:

\begin{equation}
\left[ \begin{array}{cc} \bf{\hat{u}^{(1)}} \\ \bf{\hat{u}^{(2)}} \end{array} \right]  = ((\bf{{\Sigma}_g} \otimes \mathbf{I_{p}})^{-1} + \bf{G'}\bf{{\Sigma}_e^{-1}}\bf{G})^{-1}  \bf{G'}\bf{{\Sigma}_e^{-1}}  \left[ \begin{array}{cc} \bf{{y}^{(1)}} \\ \bf{{y}^{(2)}} \end{array} \right] 
\end{equation}
where $\bf{G} = \left[ \begin{array}{cc} \bf{G^{(1)}} & \bf{0} \\ \bf{0} & \bf{G^{(2)}} \end{array} \right]$ and $\otimes$ stands for the Kronecker product. After some rearrangements, we can see that this is essentially equivalent to the solution to a bivariate ridge regression given a certain set of regularization parameters:
\begin{equation}
\left[ \begin{array}{cc} \bf{\hat{u}^{(1)}} \\ \bf{\hat{u}^{(2)}} \end{array} \right]  = ( \bf{K} \otimes \mathbf{I_{p}} + \bf{G'} \bf{G})^{-1}  \bf{G'} \left[ \begin{array}{cc} \bf{{y}^{(1)}} \\ \bf{{y}^{(2)}} \end{array} \right] 
\end{equation}
where ${\mathbf K} = (1-\rho_g^2)^{-1}\left[ \begin{array}{cc} {\sigma^2_{e_1}}/{\sigma^2_{g_1}} & -\rho_g {\sigma^2_{e_2}}/({\sigma_{g_1} \sigma_{g_2}}) \\ -\rho_g {\sigma^2_{e_1}}/({\sigma_{g_1} \sigma_{g_2}}) & {\sigma^2_{e_2}}/{\sigma^2_{g_2}} \end{array} \right]$ is the ``ridge matrix'' as defined in Brown and Zidek\cite{brown1980adaptive}. \\ 

We note that directly calculating Eq. (8) is infeasible because it involves inversion of a $2p$ by $2p$ matrix where $p$ is total number of SNPs, which could be in the order of $10^6$ in a typical GWAS. Therefore we used the matrix inversion lemma to rearrange it as: 

\begin{equation}
\left[ \begin{array}{cc} \bf{\hat{u}^{(1)}} \\ \bf{\hat{u}^{(2)}} \end{array} \right]  = (\mathbf{K^{-1}} \otimes \mathbf{I_{p}}) \mathbf{G'} [ \mathbf{I_{(n_1 + n_2)}} + \bf{G} (\mathbf{K^{-1}} \otimes \mathbf{I_{p}}) \bf{G'} ]^{-1} \left[ \begin{array}{cc} \bf{{y}^{(1)}} \\ \bf{{y}^{(2)}} \end{array} \right] 
\end{equation}
where ${\mathbf K}^{-1} = \left[ \begin{array}{cc} {\sigma^2_{g_1}}/{\sigma^2_{e_1}} & \rho_g {\sigma_{g_1} \sigma_{g_2}}/{\sigma^2_{e_1}} \\ \rho_g {\sigma_{g_1} \sigma_{g_2}}/{\sigma^2_{e_2}} & {\sigma^2_{g_2}} / {\sigma^2_{e_2}}\end{array} \right]$. Now we only need to evaluate the inverse of a matrix of dimension $(n_1+n_2)$ by $(n_1+n_2)$, which is usually in the order of $10^3$. Given the estimates of the random effects $\bf{\hat{u}^{(1)}}$ and $\bf{\hat{u}^{(2)}}$ and two sets of validation individuals with standardized genotype matrix $\bf{G^{(1)}_{v}}$ and  $\bf{G^{(2)}_{v}}$, the predicted values for the two phenotypes are given by:
\begin{equation}
\bf{\hat{y}^{(1)}} = \bf{G^{(1)}_{v}} \bf{\hat{u}^{(1)}}, \,\, \bf{\hat{y}^{(2)}} = \bf{G^{(2)}_{v}} \bf{\hat{u}^{(2)}} 
\end{equation}

The solution of the bivariate ridge regression depends on four regularization parameters. To alleviate the computational burden of tuning the regularization parameters, we impose the constraint that $\sigma_{e_1} = \sigma_{e_2}$, which leads to a symmetric ``ridge matrix'' $\mathbf{K}$ and hence a symmetric $\mathbf{K}^{-1}$. Then we can reparameterize $\bf{K}^{-1}$ as $\left[ \begin{array}{cc} \lambda_1 & \rho_g \sqrt{\lambda_1 \lambda_2} \\ \rho_g \sqrt{\lambda_1 \lambda_2} & \lambda_2 \end{array} \right]$. As a result, we only need to tune three regularization parameters $\lambda_1$, $\lambda_2$ and $\rho_g$. $\lambda_1$ and $\lambda_2$ control the shrinkage level of genetic effects for the two phenotypes whereas $\rho_g$ controls the correlation between the genetic effects between the two phenotypes. In order to find the optimal regularization parameters, we perform grid searches with $\lambda_1$ and $\lambda_2$ ranging from a very small number $\lambda_{min}$ to a very large number ${\lambda_{max}}$ and $\rho_g \in [0, 1)$. In the real data analysis, the parameter values that give the highest cross-validation mean AUC for each phenotype were chosen. In the simulation studies, the parameter values that give the highest AUC in the validation data for each phenotype were chosen. \\

We classify the validation individuals into affected ones and unaffected ones by dichotomizing the predicted values with a grid of threshold values ranging from the largest predicted value to the smallest predicted value and obtain the \emph{receiver operating characteristic} curve by evaluating the sensitivity and specificity at each threshold value. Then we evaluate the area under \emph{receiver operating characteristic} curve (AUC) as a measure of the prediction accuracy. \\

In the bivariate ridge regression, we used cross-validation to tune the regularization parameters instead of trying to estimate the parameters from the data as in \cite{lee2012estimation}. There are primarily two reasons: 1) the AI-REML algorithm that Lee et al. used in \cite{lee2012estimation} sometimes fails to converge, for example on the CD-UC data set. 2) The estimated parameters are based on normality assumptions and henceforth may not be optimal for predicting binary phenotypes. \\

\subsection*{Simulation Study}
In order to examine the relationship between the gain of predictive power and the level of genetic correlation between two diseases, we performed the following simulation studies based on the classical liability threshold model \cite{falconer1965inheritance} to simulate the case-control data. Specifically, given a desired sample size $N$, the total number of SNPs $p$, the proportion of cases in the case-control data $P$ and the disease prevalence $Q$, genotypes of a cohort of at least $NP/Q$ individuals were generated as follows: first, the MAF of $p$ SNPs were uniformly drawn from $[0.05, 0.5]$; then the genotypes (the number of minor allele copies) for the $NP/Q$ individuals at each SNP were drawn from a binomial distribution of size two and the probability of success being the MAF of the corresponding SNP. \\

After the genotypes were generated, $m$ casual SNPs were randomly chosen with each causal SNP carrying a per-minor-allele effect drawn from a normal distribution with zero mean and variance of $\frac{h^2}{(1-h^2)f_j(1-f_j)m}$ where $h^2$ is the desired level of variance explained by all SNPs on the liability scale and $f_j$ is the MAF of the corresponding causal SNP. Then the environmental effect on the liability scale for each individual was independently drawn from a standard normal distribution (zero mean and unit variance). Then, we obtained the liability for each individual by adding up the genetic effects conferred by all causal SNPs and environmental effect. Once the liabilities were obtained, individuals with liabilities greater than the $1-Q$ quantile were classified as cases and the others classified as controls. Then $NP$ cases and $N(1-P)$ controls were randomly drawn from the cohort. When simulating data for two diseases simultaneously, the two data set were simulated from two disjoint cohorts. The MAF for each SNP was the same for the two cohorts. A total of $m'$ causal SNPs were chosen to be shared between the two diseases to mimic the shared genetic basis between them. We assume shared causal SNPs have a correlation of 0.8 between their liability scale effect sizes in the two diseases. \\

In our simulation studies, we set the sample size $N$ at 1,000 or 2,000. The number of causal SNPs $m$ was chosen to be 1,000 or 2,000. $P$ was fixed at 0.5 and $K$ was fixed at 0.05. We assumed the two diseases had the same number of casual SNPs. We varied the proportion of the causal SNPs that were shared between the two diseases $\gamma = m'/m$ from $0$ to $1$ to model different levels of genetic correlation and investigated the gain of predictive power under different settings. We also investigated the effects of unequal sample size and unequal $h^2$ between the two diseases. \\

\section{Results}\label{sec_results}

\subsection*{Results from Real Data }
We analyzed GWAS data for bipolar and related disorders (BARD) and schizophrenia (SZ) that we downloaded from the dbGaP database (\burlalt{{https://dbgap.ncbi.nlm.nih.gov}}{https://dbgap.ncbi.nlm.nih.gov}). After pre-processing (see Methods), the combined data set consisted of 3,226 individuals (653 cases for BARD, 1,170 cases for SZ and 1,403 controls) with genotypes at 298,604 SNPs. In order to organize two case-control data sets for the two diseases, we randomly partitioned the controls into two disjoint sets of individuals and assigned each set to one of the diseases. We partitioned the control individuals so that the proportions of cases for the two data sets were approximately equal, i.e. we assigned 505 controls to BARD and 898 controls to SZ. The random partitioning was repeated 50 times. \\
	
We then used the bivariate ridge regression method \cite{brown1980adaptive} described above to predict jointly the disease status for BARD and SZ using this data set. Prediction performance was evaluated through five-fold cross-validation, i.e., the data were randomly partitioned into five equal-sized folds and four of them were used as the training set and the other one was used as the validation set each time. We calculated the mean AUC (area under the \emph{receiver-operating-characteristic} curve) for the five folds to measure prediction performance. For comparison, we also evaluated prediction performance of three other prediction methods: univariate ridge regression, SVM (support vector machine) \cite{fan2008liblinear} with linear kernel, and LASSO (least absolute shrinkage and selection operator) \cite{tibshirani1996regression} that treat each disease separately (Figure 1). Note that the bivariate ridge regression method we used is a direct generalization of the univariate ridge regression (see Methods). Therefore the utility of pleiotropy can be best demonstrated through the comparison between the two approaches. Indeed, the bivariate ridge regression method performed better than the univariate ridge regression for both BARD and SZ. We also note that the gain of AUC of BARD (0.041) was much larger than that of SZ (only 0.013), probably as a result of the sample size difference between the two diseases. SVM also performed better than univariate ridge regression, but it was still outperformed by bivariate ridge regression in BARD (with an AUC of 0.031) although in SZ it performed better than bivariate ridge regression by 0.016. LASSO, which prefers a sparse genetic architecture with a small number of major (strong effect) SNPs, did not perform well for both diseases. \\

\begin{figure}[!h]
  \includegraphics[width=\textwidth]{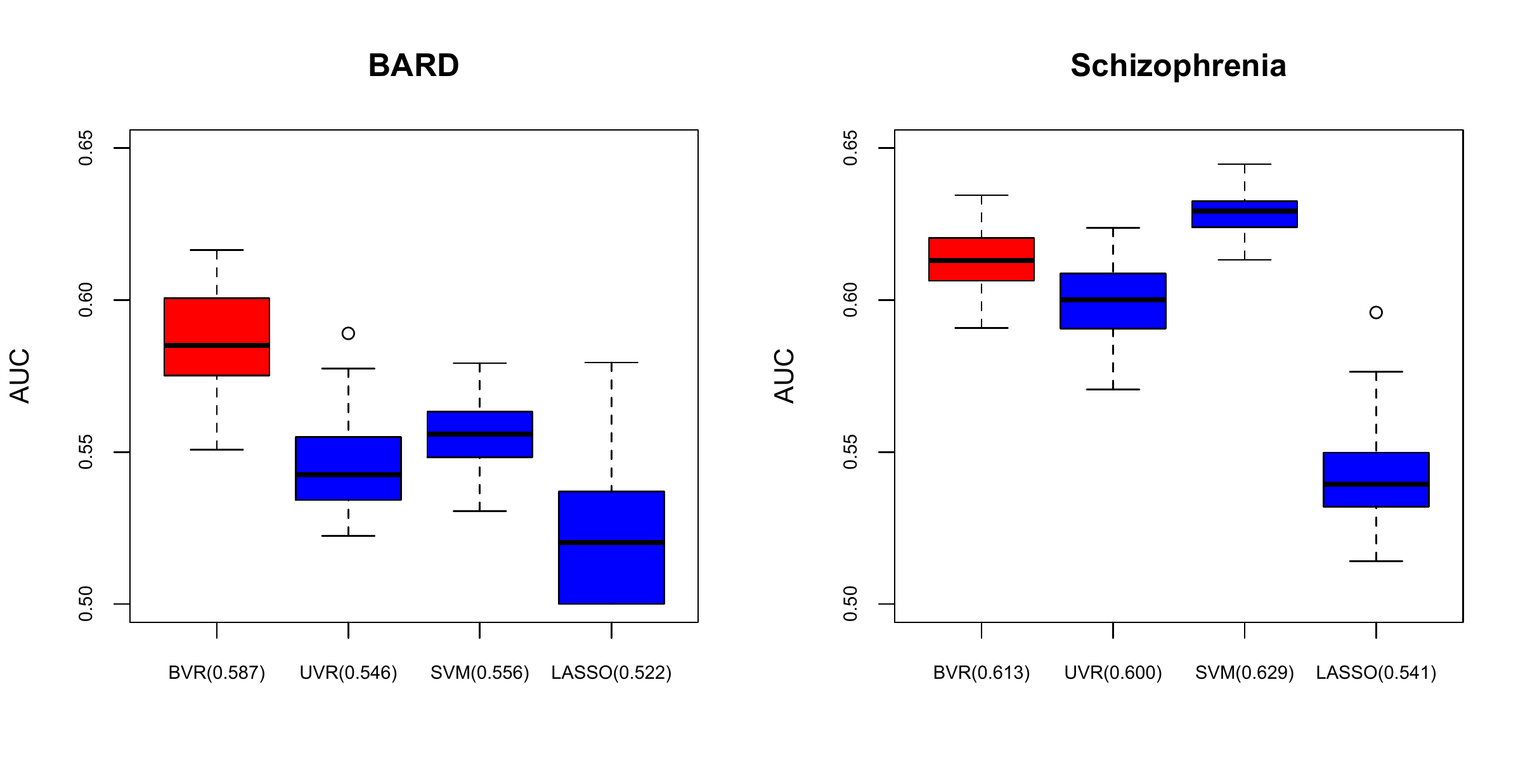}
\caption{Prediction accuracy of different methods on the BARD-SZ data. ``BVR'' and ``UVR'' stand for bivariate ridge regression and univariate ridge regression respectively. The numbers in the brackets are the mean AUCs achieved by each method in the 50 repeats.}
\label{fig:1}       
\end{figure}

To evaluate whether the improvement was the result of the genetic correlation between BARD and SZ or an artificial effect of the bivariate ridge regression method, we randomly selected 25\%, 50\%, 75\% and all of the SNPs and shuffled their identities in the BARD samples while keeping the SNP identities unchanged in the SZ samples. This led to reduced genetic correlation between BARD and SZ and consequently reduced improvement of prediction accuracy for the bivariate ridge regression (Figure 2). In particular, when all the SNPs were shuffled, the performance of bivariate ridge regression was almost the same as univariate ridge regression, confirming that the gain of prediction accuracy of the bivariate ridge regression indeed came from specific SNP, and thus the genetic correlation between the two phenotypes. \\

\begin{figure}[!h]
  \includegraphics[width=\textwidth]{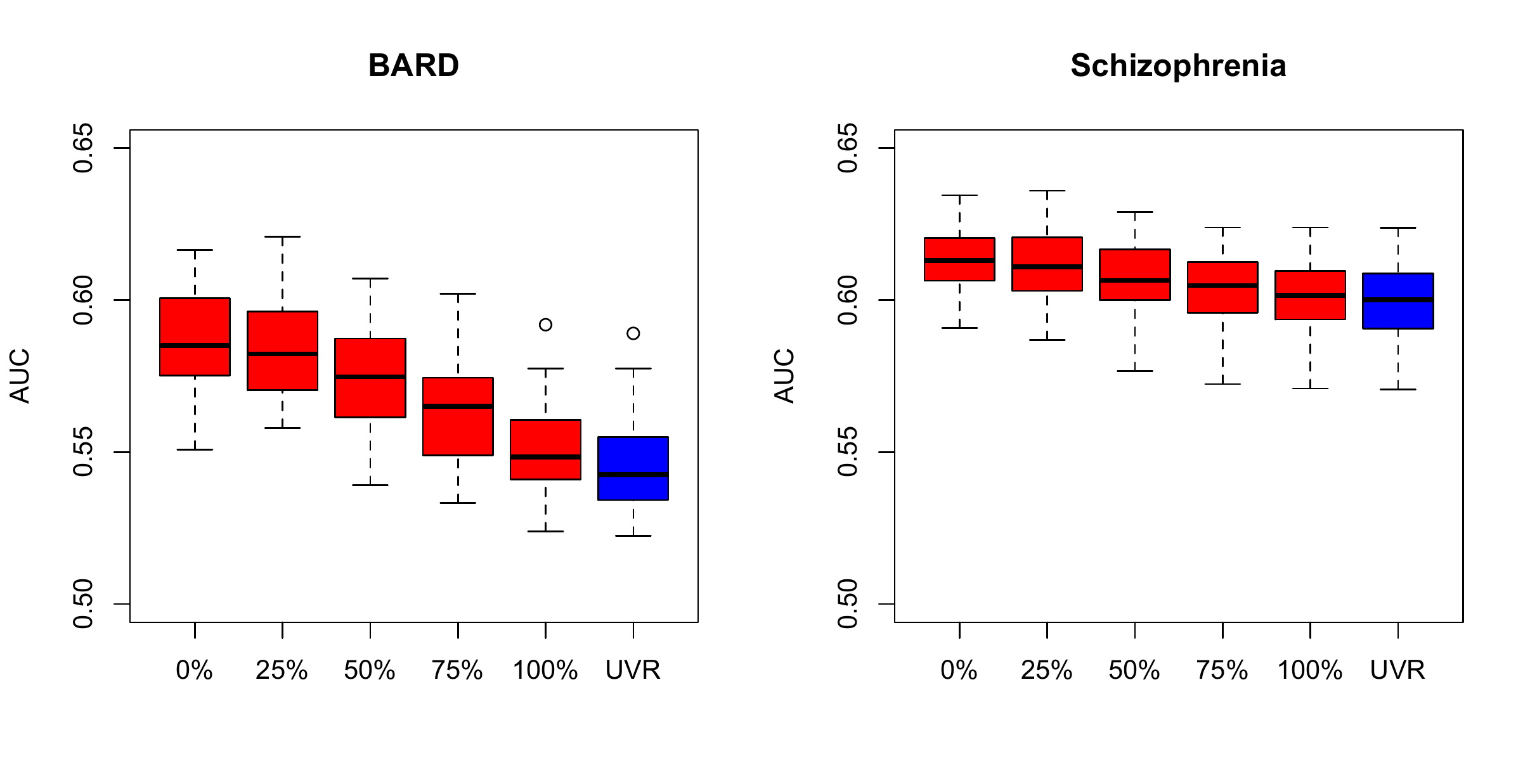}
\caption{Prediction accuracy of the bivariate ridge regression after shuffling the SNP identities and of ridge regression. Red plots represent the results of bivariate ridge regression and blue plots represent those of ridge regression. The percentage below each red plot represents the fraction SNPs that were shuffled.}
\label{fig:2}       
\end{figure}

We also did the same analysis on a GWAS data set for Crohn's disease (CD) and a GWAS data set for ulcerative colitis (CD) obtained from the dbGap database. The CD data set consists of 515 unaffected subjects and 513 affected subjects with non-Jewish European ancestry and the UC data set consists of 902 affected subjects with non-Jewish European ancestry. Because there are no control samples for the UC data set, we obtained 902 samples from the Illumina iControl database (\burlalt{{http://www.illumina.com/science/icontroldb.ilmn}}{http://www.illumina.com/science/icontroldb.ilmn}) that were genetically matched with the UC samples to minimize the confounding effect of potential population stratification (see Methods). After pre-processing (see Methods), we obtained a data set consisting of 513 affected subjects for CD, 515 unaffected subjects for CD, 902 affected subjects for UC and 902 unaffected subjects for UC with genotypes for 241,649 SNPs. As opposed to the BARD and SZ data set, the control samples for CD and UC were fixed instead of being randomly partitioned.\\

Prediction accuracies of different methods on the CD-UC data set were also evaluated through five-fold cross-validation, with the cross-validation repeated 50 times. The results are shown in Figure 3. Similar to the results for the BARD-SZ data, bivariate ridge regression performed better than univariate ridge regression in both diseases. Moreover, the increase of AUC for CD, which has a smaller sample size, was larger than that of UC (0.055 versus 0.019). SVM also outperformed univariate ridge regression in both diseases but was outperformed by bivariate ridge regression in CD. \\

\begin{figure}[!h]
  \includegraphics[width=\textwidth]{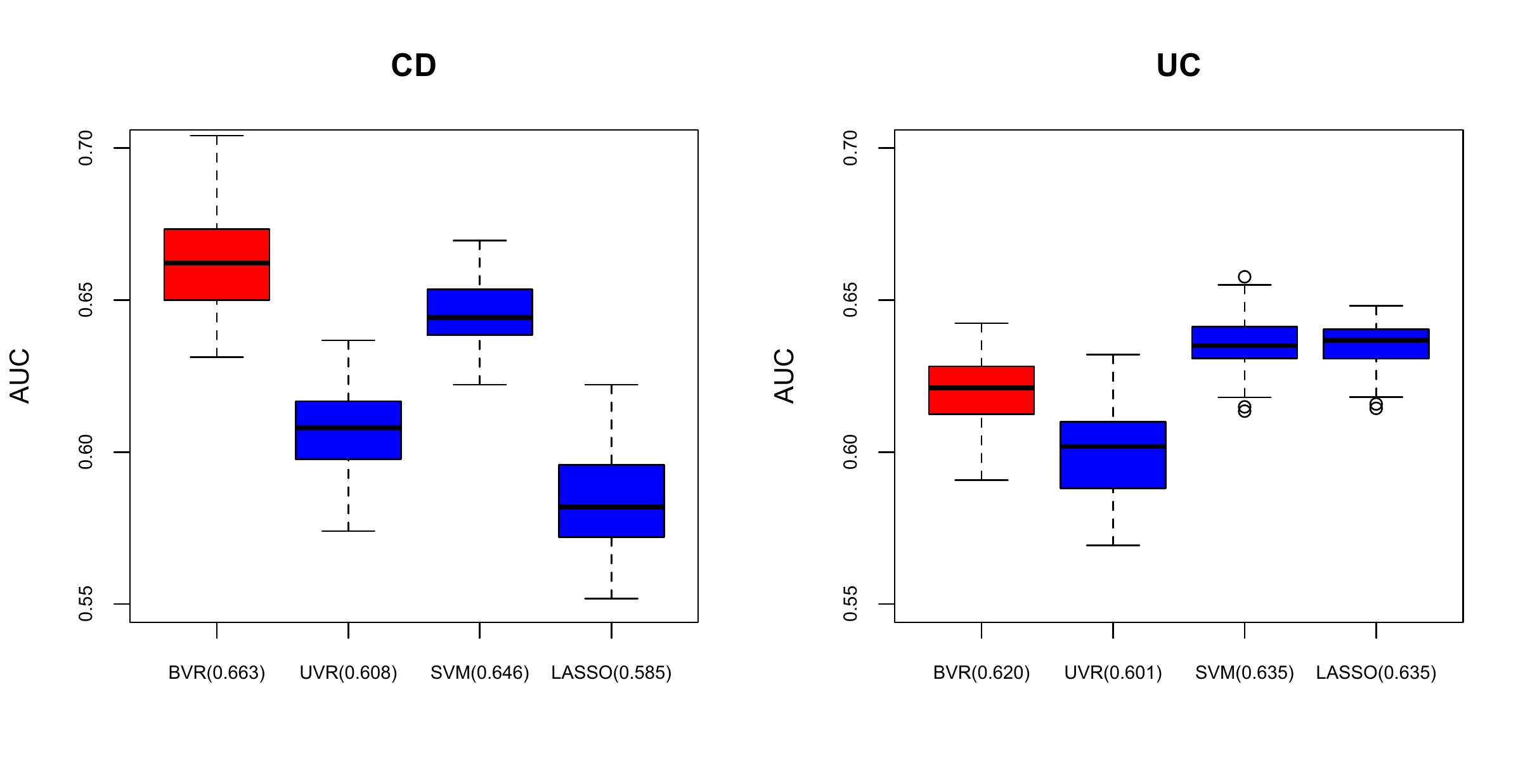}
\caption{Prediction accuracy of different methods on the CD-UC data. ``BVR'' and ``UVR'' are defined as in Figure 1. The numbers in the brackets are the mean AUCs achieved by each method in the 50 repeats. }
\label{fig:3}       
\end{figure}

Similar to the BARD-SZ data set, we performed random shuffling of the identities of the SNPs in the CD samples while keeping their identities in the UC samples. The results are shown in Figure 4. As expected, the improvement of prediction accuracy of bivariate ridge regression also decreases as the fraction of SNPs increases and eventually diminishes when all SNPs are shuffled.\\

\begin{figure}[!h]
  \includegraphics[width=\textwidth]{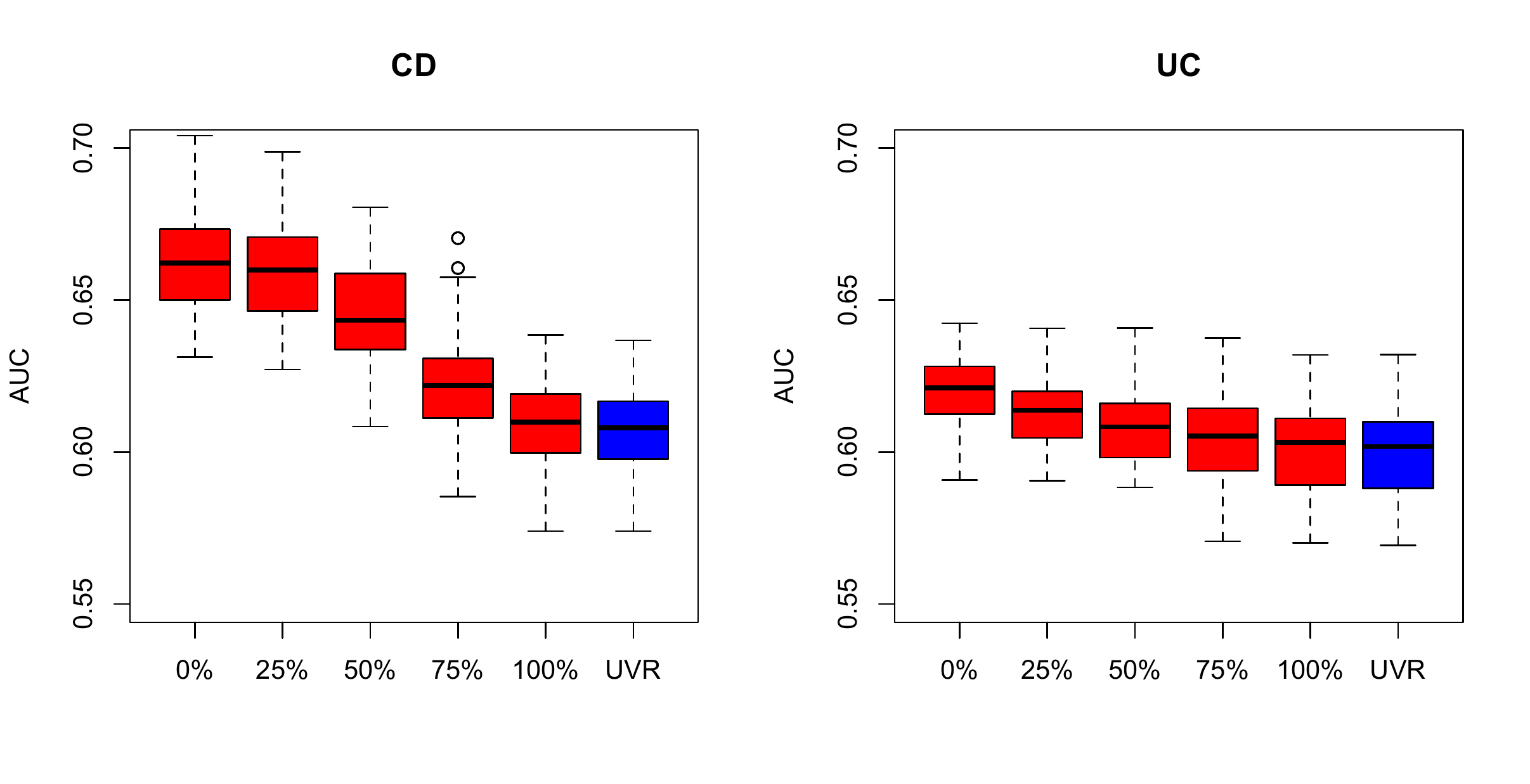}
\caption{Prediction accuracy of the bivariate ridge regression after shuffling the SNP identities and of ridge regression. Red plots represent the results of bivariate ridge regression and blue plots represent those of the ridge regression. The percentage below each red plot represents the fraction SNPs that were shuffled.}
\label{fig:4}       
\end{figure}
 
 \subsection*{Results from Simulation Study}
We simulated data sets for two genetically correlated diseases to evaluate the utility of genetic correlation in genetic risk prediction. We simulated a total of $p = 20,000$ SNPs. We let the sample size $N$ to be 1000 or 2000. The number of causal SNPs $m$ was also chosen to be 1000 or 2000. We assumed the two traits had the same number of casual SNPs. We varied the proportion of the causal SNPs that were shared between the two diseases $\gamma$ from $0$ to $1$ to mimic different levels of genetic correlation. We assumed that the shared causal SNPs had a correlation of $0.8$ between their effect sizes on the two diseases. Our empirical results suggest that the realized genetic correlation is approximately $0.8\gamma$ under this setup (results not shown). We simulated two $h^2$ levels (on liability scale, see \cite{lee2011estimating}): $0.3$ and $0.6$. Disease prevalence was set to be $0.05$ and equal numbers of cases and controls were drawn from the simulated population. The simulations were repeated for 25 times in each scenario. In each repeat, a training set and a validation set were independently generated and were used to evaluate the prediction accuracies of bivariate ridge regression and univariate ridge regression. \\

We first considered the scenario where both diseases had the same number of samples and the same $h^2$ level. The results for $m = 1000$ are shown in Figure 5 and the result for $m = 2000$ are shown in Figure S1. As expected, when the genetic correlation between the two diseases was zero, the prediction accuracy of bivariate ridge regression was almost the same as univariate ridge regression. As the genetic correlation increased, the AUC for bivariate ridge regression also increased and the improvement over univariate ridge regression became quite noticeable as long as $\gamma \geq 0.5$, which corresponds to a genetic correlation of approximately 0.4. \\

\begin{figure}[!h]
  \includegraphics[width=\textwidth]{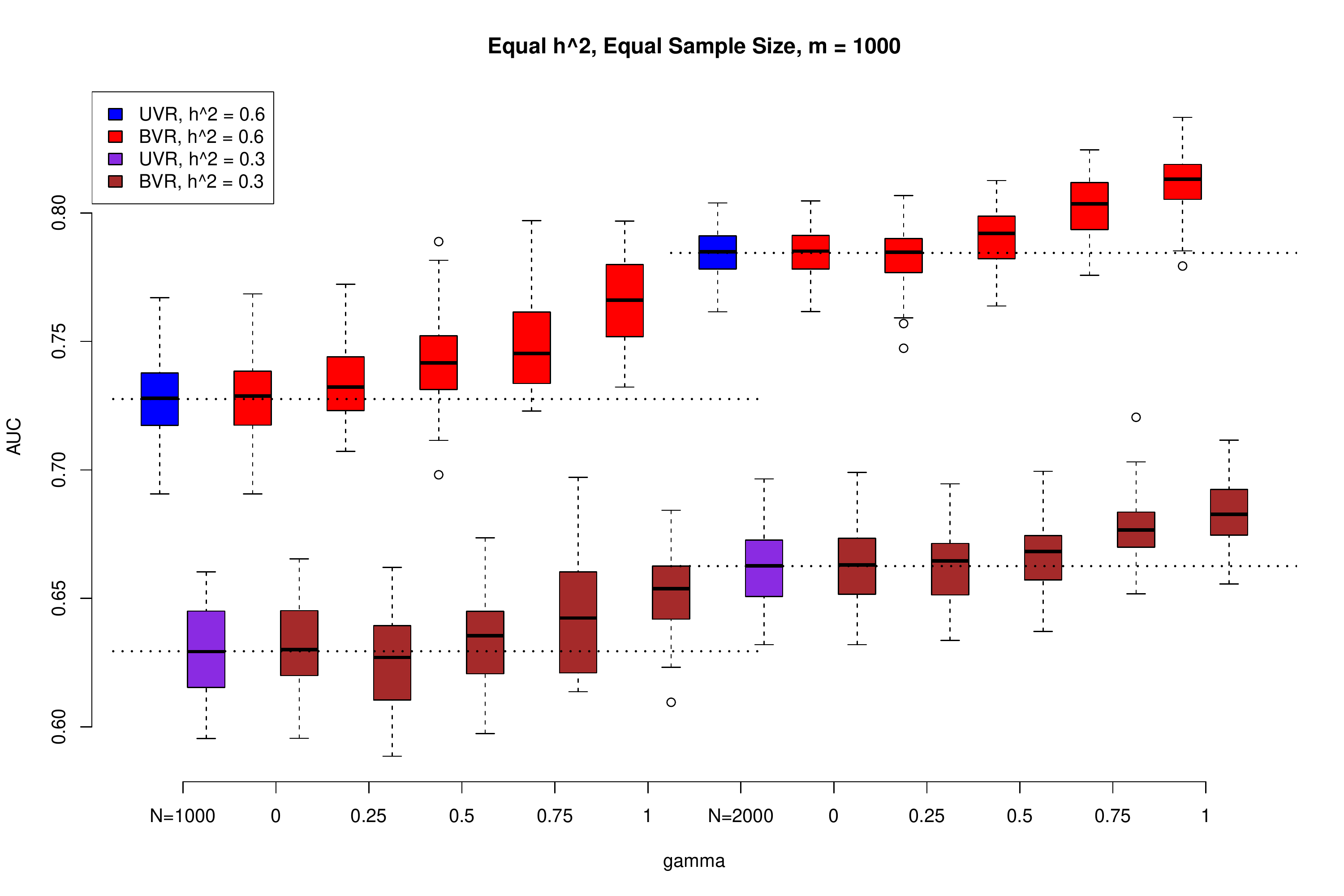}
\caption{Simulation results for the case when the two diseases have equal sample sizes and $h^2$ levels and $m = 1000$. ``BVR'' and ``UVR'' are defined as in Figure 1. Two $h^2$ levels ($0.3$ and $0.6$) and two sample sizes ($1000$ and $2000$) were simulated. The proportion of shared causal SNPs, $\gamma$ was varied from $0$ to $1$ with an increment of $0.25$. The numbers below the UVR box plots are the sample sizes. Following the UVR box plots are the box plots representing the results of BVR with the same sample sizes at different $\gamma$ values (below the BVR box plots).}
\label{fig:5}       
\end{figure}

We also examined the case when the sample sizes (Figure 6 and Figure S2) or the $h^2$ levels (Figure 7 and Figure S3) were different between the two traits. When the sample sizes were different but the  $h^2$ levels were the same between the two diseases, the prediction accuracy of the disease with the smaller sample size improved more than that of the disease with the larger sample size by joint modeling, which echoes with the results from the real data analysis.  \\

\begin{figure}[!h]
  \includegraphics[width=\textwidth]{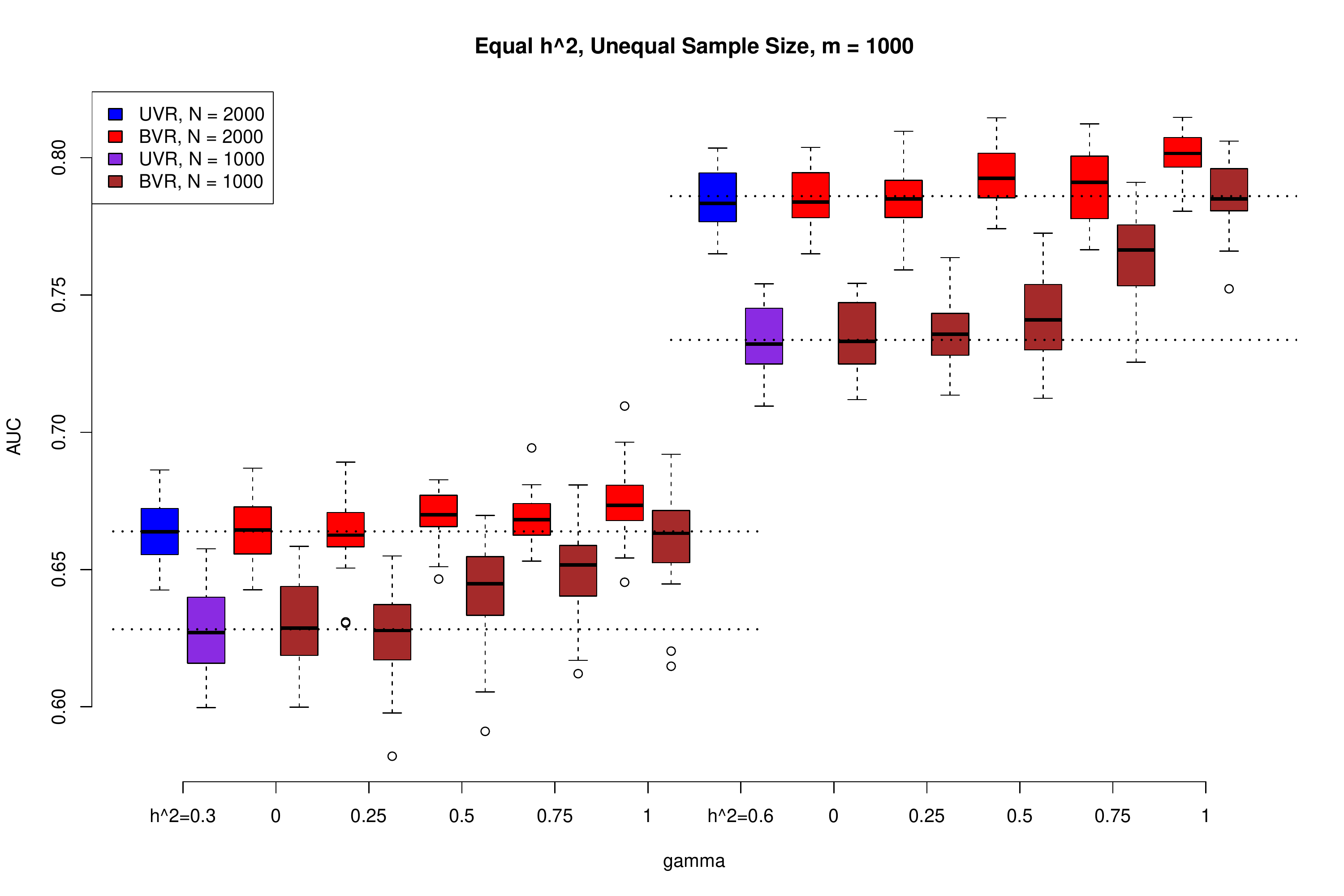}
\caption{Simulation results for the case when the two diseases have unequal sample sizes and equal $h^2$ levels and $m = 1000$. ``BVR'' and ``UVR'' are defined as in Figure 1. One of the diseases has 2000 samples and the other has 1000 samples. Two $h^2$ levels ($0.6$ and $0.3$) were simulated. The proportion of shared causal SNPs, $\gamma$ was varied from $0$ to $1$ with an increment of $0.25$. The numbers below the UVR box plots are the $h^2$ levels. Following the UVR box plots are the results of BVR with the same $h^2$ levels at different $\gamma$ values (below the BVR box plots).}
\label{fig:6}       
\end{figure}

\begin{figure}[!h]
  \includegraphics[width=\textwidth]{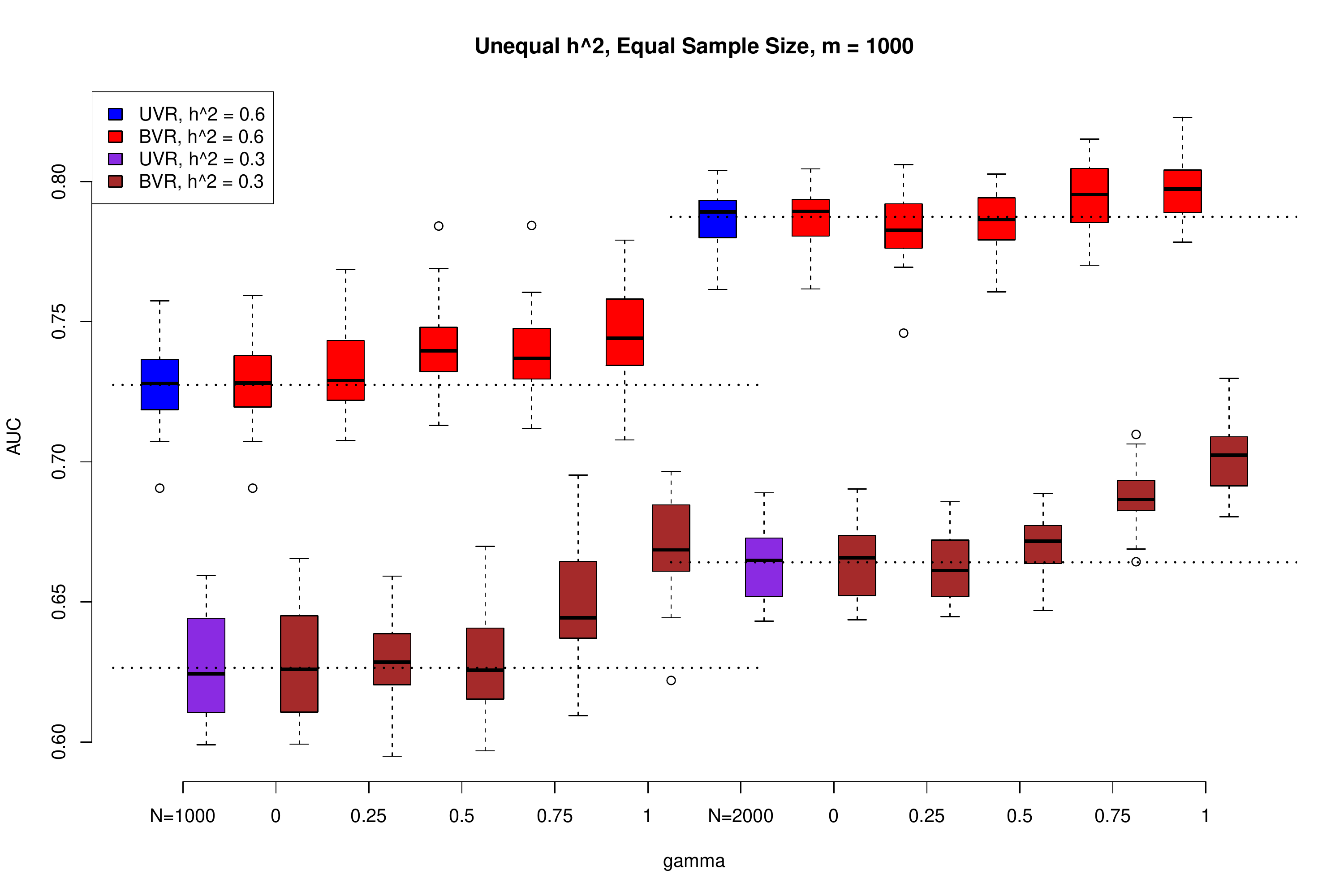}
\caption{Simulation results for the case when the two diseases have equal sample sizes and unequal $h^2$ levels and $m = 1000$. ``BVR'' and ``UVR'' are defined as in Figure 1. One of the diseases has $h^2 = 0.6$ and the other has $h^2 = 0.3$. Two sample sizes ($2000$ and $1000$) were simulated. The proportion of shared causal SNPs, $\gamma$ was varied from $0$ to $1$ with an increment of $0.25$. The numbers below the UVR box plots are the sample sizes. Following the UVR box plots are the results of BVR with the same sample sizes at different $\gamma$ values (below the BVR box plots).}
\label{fig:7}       
\end{figure}

\section{Discussion}\label{sec_discussions}

Genetic risk prediction is a challenging task due to the existence of numerous weak-effect variants and is also bottlenecked by the available sample size of GWAS. A promising yet under-appreciated strategy to increase the effective training sample size and thus to improve prediction accuracy is to integrate diseases that are genetically correlated. In this paper we performed a systematic investigation of the utility of genetic correlation in genetic risk prediction using a bivariate ridge regression method. In the real data analysis, we demonstrated that gain of predictive power can be achieved by using the genetic correlation between phenotypes. In the simulation study, the results confirmed our findings in the real data, as well as offered us insights into the relationship between gain of prediction accuracy and the level of genetic correlation between two phenotypes. These results can provide important guide for researchers to make productive use of pleiotropy while developing disease risk prediction models in the future.\\

In the real data and simulation studies, we found that the prediction accuracy of the disease with the smaller sample size benefited more from the joint modeling than that of the disease with the larger sample size. The possible reason is that when jointly modeling the two data sets, the larger data set contains more information or statistical evidence about the genetic effects shared by the two diseases than the smaller data set, and therefore more information can be borrowed from the larger data set to the smaller data set than the opposite way. In the simulation studies, we also observed that when the sample sizes were the same but the  $h^2$ levels were different, joint modeling offered more benefit to the low-heritability disease than that of the high-heritability disease. A similar explanation can be made: the high-heritability disease data set can provide stronger statistical evidence about the genetic effects shared by the two diseases than the low-heritability disease. Note that the two explanations are based on the simulation setup that the number of associated SNPs is the same for the two diseases, all the SNPs are independent, and the genetic effects are randomly distributed among the causal SNPs.\\

To our knowledge, only Hartley et al \cite{hartley2012bayesian,hartley2013pleiogrip} have studied pleiotropy in genetic risk prediction to date. However, our study is distinct from theirs in two aspects. Firstly, Hartley et al. focused on methodological development whereas our study emphasizes the utility of genetic correlation using real GWAS data sets. In fact, our study is the first one to demonstrate a substantial contribution of genetic correlation to prediction accuracy on real data. Secondly, their method requires that all the phenotypes of interest are observed on all individuals, which limits its applications. In contrast, we consider the case where two phenotypes are observed on two disjoint sets of individuals. Not only does this allow more GWAS data sets to be integrated, we can expect much more improvement of the predictive ability since the sample size is effectively increased by combining the data for the two phenotypes whereas there is no increase in the sample size in the scenario considered by Hartley and colleagues. In addition, even when we have phenotype information from multiple traits for each study subject, the bivariate mixed effect model may not appropriately accommodate specific, sometimes complex, ascertainment schemes used to collect the GWAS samples. This may lead to potential biases and reduced prediction accuracy, e.g. causing some shared genetic effects between the two traits to be cancelled out. \\

Solovieff et al \cite{solovieff2013pleiotropy} pointed out several sources of spurious pleiotropy. One is sample ascertainment bias. In our real data examples, each ``case'' individual is affected by only one disease and the controls are not shared by the two diseases. Therefore, it is unlikely that our study is affected by ascertainment bias. Another source is misdiagnosis, e.g. people with schizophrenia may be misclassified as bipolar disorder. However they also noted that the misclassification rate has to be very high to generate a substantial genetic correlation, which suggests the gain of prediction accuracy in our results may still primarily be a result of the true genetic correlation between the two diseases.\\

The method that we used to jointly predict two phenotypes is a bivariate generalization to the ridge regression. Although the comparison between bivariate ridge regression and univariate suggests that pleiotropy can indeed substantially contribute to genetic risk prediction, we do note that there is also a need to develop better methods for joint prediction of multiple phenotypes. For example, SVM always does better than univariate ridge regression. Development of such methods that jointly predict multiple phenotypes is an important future task. In this study, we demonstrated the utility of pleiotropy through two pairs of diseases that are known to share a lot of common genetic bases. Another important future task is to comprehensively examine the genetic correlations between other complex human diseases and their utility in risk prediction.\\

\begin{acknowledgements}

This study was supported by NIH grants R01 AA11330, AA017535, DA030976, GM59507, VA Cooperative Studies Program 572, and a fellowship from China Scholarship Council. We also thank Yale University High Performance Computing Center (funded by NIH RR19895) for the computation resource and data storage. \\

Funding support for the Whole Genome Association Study of Bipolar Disorder was provided by the National Institute of Mental Health (NIMH) and the genotyping of samples was provided through the Genetic Association Information Network (GAIN). The datasets used for the analyses described in this manuscript were obtained from the database of Genotypes and Phenotypes (dbGaP) found at http://www.ncbi.nlm.nih.gov/gap through dbGaP accession number phs000017.v3.p1. Samples and associated phenotype data for the Collaborative Genomic Study of Bipolar Disorder were provided by the The NIMH Genetics Initiative for Bipolar Disorder. Data and biomaterials were collected in four projects that participated in NIMH Bipolar Disorder Genetics Initiative. From 1991-98, the Principal Investigators and Co-Investigators were: Indiana University, Indianapolis, IN, U01 MH46282, John Nurnberger, M.D., Ph.D., Marvin Miller, M.D., and Elizabeth Bowman, M.D.; Washington University, St. Louis, MO, U01 MH46280, Theodore Reich, M.D., Allison Goate, Ph.D., and John Rice, Ph.D.; Johns Hopkins University, Baltimore, MD U01 MH46274, J. Raymond DePaulo, Jr., M.D., Sylvia Simpson, M.D., MPH, and Colin Stine, Ph.D.; NIMH Intramural Research Program, Clinical Neurogenetics Branch, Bethesda, MD, Elliot Gershon, M.D., Diane Kazuba, B.A., and Elizabeth Maxwell, M.S.W. Data and biomaterials were collected as part of ten projects that participated in the NIMH Bipolar Disorder Genetics Initiative. From 1999-03, the Principal Investigators and Co-Investigators were: Indiana University, Indianapolis, IN, R01 MH59545, John Nurnberger, M.D., Ph.D., Marvin J. Miller, M.D., Elizabeth S. Bowman, M.D., N. Leela Rau, M.D., P. Ryan Moe, M.D., Nalini Samavedy, M.D., Rif El-Mallakh, M.D. (at University of Louisville), Husseini Manji, M.D. (at Wayne State University), Debra A. Glitz, M.D. (at Wayne State University), Eric T. Meyer, M.S., Carrie Smiley, R.N., Tatiana Foroud, Ph.D., Leah Flury, M.S., Danielle M. Dick, Ph.D., Howard Edenberg, Ph.D.; Washington University, St. Louis, MO, R01 MH059534, John Rice, Ph.D, Theodore Reich, M.D., Allison Goate, Ph.D., Laura Bierut, M.D. ; Johns Hopkins University, Baltimore, MD, R01 MH59533, Melvin McInnis M.D. , J. Raymond DePaulo, Jr., M.D., Dean F. MacKinnon, M.D., Francis M. Mondimore, M.D., James B. Potash, M.D., Peter P. Zandi, Ph.D, Dimitrios Avramopoulos, and Jennifer Payne; University of Pennsylvania, PA, R01 MH59553, Wade Berrettini M.D.,Ph.D. ; University of California at Irvine, CA, R01 MH60068, William Byerley M.D., and Mark Vawter M.D. ; University of Iowa, IA, R01 MH059548, William Coryell M.D. , and Raymond Crowe M.D. ; University of Chicago, IL, R01 MH59535, Elliot Gershon, M.D., Judith Badner Ph.D. , Francis McMahon M.D. , Chunyu Liu Ph.D., Alan Sanders M.D., Maria Caserta, Steven Dinwiddie M.D., Tu Nguyen, Donna Harakal; University of California at San Diego, CA, R01 MH59567, John Kelsoe, M.D., Rebecca McKinney, B.A.; Rush University, IL, R01 MH059556, William Scheftner M.D. , Howard M. Kravitz, D.O., M.P.H., Diana Marta, B.S., Annette Vaughn-Brown, MSN, RN, and Laurie Bederow, MA; NIMH Intramural Research Program, Bethesda, MD, 1Z01MH002810-01, Francis J. McMahon, M.D., Layla Kassem, PsyD, Sevilla Detera-Wadleigh, Ph.D, Lisa Austin,Ph.D, Dennis L. Murphy, M.D.\\

Funding support for the Genome-Wide Association of Schizophrenia Study was provided by the National Institute of Mental Health (R01 MH67257, R01 MH59588, R01 MH59571, R01 MH59565, R01 MH59587, R01 MH60870, R01 MH59566, R01 MH59586, R01 MH61675, R01 MH60879, R01 MH81800, U01 MH46276, U01 MH46289 U01 MH46318, U01 MH79469, and U01 MH79470) and the genotyping of samples was provided through the Genetic Association Information Network (GAIN). The datasets used for the analyses described in this manuscript were obtained from the database of Genotypes and Phenotypes (dbGaP) found at http://www.ncbi.nlm.nih.gov/gap through dbGaP accession number phs000021.v3.p2. Samples and associated phenotype data for the Genome-Wide Association of Schizophrenia Study were provided by the Molecular Genetics of Schizophrenia Collaboration (PI: Pablo V. Gejman, Evanston Northwestern Healthcare (ENH) and Northwestern University, Evanston, IL, USA). \\

The NIDDK IBD Genetics Consortium Crohn's Disease Genome-Wide Association Study was conducted by the NIDDK IBD Genetics Consortium Crohn's Disease Genome-Wide Association Study Investigators and supported by the National Institute of Diabetes and Digestive and Kidney Diseases (NIDDK). The data and samples from the NIDDK IBD Genetics Consortium Crohn's Disease Genome-Wide Association Study reported here were supplied by the NIDDK Central Repositories. The datasets used for the analyses described in this manuscript were obtained from the database of Genotypes and Phenotypes (dbGaP) found at http://www.ncbi.nlm.nih.gov/gap through dbGaP accession number phs000130.v1.p1. This manuscript was not prepared in collaboration with Investigators of the NIDDK IBD Genetics Consortium Crohn's Disease Genome-Wide Association Study and does not necessarily reflect the opinions or views of the NIDDK IBD Genetics Consortium Crohn's Disease Genome-Wide Association Study, the NIDDK Central Repositories, or the NIDDK. \\

The NIDDK IBD Genetics Consortium Ulcerative Colitis Genome-Wide Association Study was conducted by the NIDDK IBD Genetics Consortium Ulcerative Colitis Genome-Wide Association Study Investigators and supported by the National Institute of Diabetes and Digestive and Kidney Diseases (NIDDK). The data and samples from the NIDDK IBD Genetics Consortium Ulcerative Colitis Genome-Wide Association Study reported here were supplied by the NIDDK Central Repositories. The datasets used for the analyses described in this manuscript were obtained from the database of Genotypes and Phenotypes (dbGaP) found at http://www.ncbi.nlm.nih.gov/gap through dbGaP accession number phs000345.v1.p1. This manuscript was not prepared in collaboration with Investigators of the NIDDK IBD Genetics Consortium Ulcerative Colitis Genome-Wide Association Study and does not necessarily reflect the opinions or views of the NIDDK IBD Genetics Consortium Ulcerative Colitis Genome-Wide Association Study, the NIDDK Central Repositories, or the NIDDK. \\
\end{acknowledgements}


%


The authors declare that they have no conflict of interest.

\bibliographystyle{spmpsci}      
\bibliography{ref}   


\end{document}